
\\




\font\seventeenss=cmss17

\font\fourteenrm=cmr12 scaled \magstep1
\font\twelverm=cmr12
\font\tenrm=cmr10
\font\eightrm=cmr8
\font\twelvebf=cmbx12
\font\twelveit=cmti12
%
\font\twelvemi=cmmi12
\font\ninemi=cmmi9
\font\sevenmi=cmmi7
%
\font\twelvesy=cmsy10 scaled \magstep1
\font\ninesy=cmsy9
\font\sevensy=cmsy7
%
\font\twelveex=cmex10 scaled \magstep1
%
%
\def\twelvepoint{
  \textfont0=\twelverm \scriptfont0=\tenrm  \scriptscriptfont0=\eightrm
  \textfont1=\twelvemi \scriptfont1=\ninemi \scriptscriptfont1=\sevenmi
  \textfont2=\twelvesy \scriptfont2=\ninesy \scriptscriptfont2=\sevensy
  \textfont3=\twelveex \scriptfont3=\twelveex  \scriptscriptfont3=\twelveex
  \textfont\itfam=\twelveit
  \textfont\bffam=\twelvebf
  \normalbaselineskip=22pt
  \setbox\strutbox=\hbox{\vrule height10pt depth4.5pt width0pt}
  \def\rm{\fam0 \twelverm}
  \def\it{\fam\itfam \twelveit}
  \def\bf{\fam\bffam \twelvebf}
  \normalbaselines
}

%
\catcode`@=11 
%
\newcount\equanumber	     \equanumber=0
%
%
\newcount\chapternumber	     \chapternumber=0
\newcount\sectionnumber	     \sectionnumber=0
\newcount\equanumber	     \equanumber=0
\let\chapterlabel=0
\newtoks\chapterstyle	     \chapterstyle={\Number}
\def\chapterreset{\global\advance\chapternumber by 1
   \ifnum\equanumber<0 \else\global\equanumber=0\fi
   \sectionnumber=0 \makel@bel}
\def\makel@bel{\xdef\chapterlabel{%
\the\chapterstyle{\the\chapternumber}.}}
\def\sectionlabel{\number\sectionnumber \quad }
\def\alphabetic#1{\count255='140 \advance\count255 by #1\char\count255}
\def\Alphabetic#1{\count255='100 \advance\count255 by #1\char\count255}
\def\Roman#1{\uppercase\expandafter{\romannumeral #1}}

\def\Number#1{\number #1}
\def\unnumberedchapters{\let\makel@bel=\relax \let\chapterlabel=\relax
\let\sectionlabel=\relax \equanumber=-1 }
\def\unnumberedchapters{\let\makel@bel=\relax \let\chapterlabel=\relax
\let\sectionlabel=\relax \equanumber=-1 }
\def\chapter{
   \chapterreset
   \wlog{\string\chapter\ \chapterlabel } }
\def\section{
   \wlog{\string\section\ \chapterlabel\the\sectionnumber}
   \global\advance\sectionnumber by 1
            }
\def\Appendix{\chapternumber=0 \chapterstyle={\Alphabetic} }
%
%
%
\def\eqname#1{\relax \ifnum\equanumber<0
     \xdef#1{{\rm(\number-\equanumber)}}\global\advance\equanumber by -1
    \else \global\advance\equanumber by 1
      \xdef#1{{\rm(\chapterlabel \number\equanumber)}} \fi}
\def\eq{\eqname\?\?}

\def\eqinsert#1{\noalign{\dimen@=\prevdepth \nointerlineskip
   \setbox0=\hbox to\displaywidth{\hfil #1}
   \vbox to 0pt{\vss\hbox{$\!\box0\!$}\kern-0.5\baselineskip}
   \prevdepth=\dimen@}}
\catcode`@=12


\twelvepoint
\rm
\def\r#1{[#1]}
\def\Ref#1{\item{[#1]}}

\def\Fot#1{\item{[#1]}}
\def\NPBsupl#1#2#3{Nucl. Phys. {\bf B} (Proc. Suppl.) {\bf #1} (#2) #3}
\def\NPB#1#2#3{Nucl. Phys. {\bf B #1} (#2) #3}
\def\PLB#1#2#3{Phys. Lett.  {\bf B #1} (#2) #3}
\def\PRD#1#2#3{Phys. Rev. {\bf D #1} (#2) #3}
\def\PRL#1#2#3{Phys. Rev. Lett. {\bf  #1} (#2) #3}
\def\PROG#1#2#3{Prog. Theor. Phys. {\bf #1} (#2) #3}

\def\PRL#1#2#3{Phys. Rev. Lett. {\bf  #1} (#2) #3}
\def\MPLA#1#2#3{Mod. Phys. Lett. {\bf A #1} (#2) #3}
\def\Ninomiya{1} \def\ninomiya{\Ref{1}}
 \def\wilson{\Ref2}
\def\Swift{3} \def\swift{\Ref3}
 \def\smit{\Ref4}
 \def\aoki{\Ref5}
\def\Kashiwa{6} \def\kashiwa{\Ref6}
\def\Bock{7}  \def\bock{\Ref7}
  \def\alss{\Ref8}
 \def\jansen{\Ref9}
\def\BDS{10}  \def\bds{\Ref{10}}
\def\Mean{11} \def\mean{\Ref{11}}
  \def\ivsd{\Ref{12}}
\def\Hop{13} \def\hop{\Ref{13}}
\def\KEhop{14}  \def\kehop{\Ref{14}}
\def\Anna{15} \def\anna{\Ref{15}}
 \def\chop{\Fot{f1}}
\def\Shift{16}  \def\shift{\Ref{16}}
\def\Rome{17} \def\rome{\Ref{17}}
\def\Ki{18} \def\ki{\Ref{18}}
\def\GPS{19}  \def\gps{\Ref{19}}
\def\Aokif{20}  \def\aokif{\Ref{20}}
  \def\qkh{\Fot{f2}}
\def\$$${\eqno\eq$$}
\def\arg#1{{\scriptstyle (} #1 {\scriptstyle )}}
\def\Lag{{\cal L}}

\def\Ln{{\rm Ln}}
\def\Tr{{\rm Tr}}

\rightline{\vbox{
                  \hbox{UTHEP-243}
                 \hbox{November 1992} } }
\vskip 20pt
\centerline{
\seventeenss
\vbox{
\hbox{ Fermion mass in the Wilson-Yukawa approach
for chiral Yukawa theory}
}}
\vskip 54pt
\centerline{
\fourteenrm
Sinya Aoki \ and \ Yoshio Kikukawa\footnote{*}{\tenrm Postdoctoral
fellow of the Japan Society for the Promotion of Science.}}
\vskip 30pt
\centerline{\it Institute of Physics, University of Tsukuba}
\centerline{\it Tsukuba, Ibaraki 305, Japan}

\vskip 36pt
\noindent
\centerline{\bf Abstract}
\par
{\leftskip 0.4 in \rightskip 0.4 in
We consider a modification of the Wilson-Yukawa model
to overcome its difficulty that the fermion mass is not proportional
to the Higgs vacuum expectation value.
In the modification scalar and fermionic regulator fields are
introduced so that all the physical fermion fields
possess shift symmetry when the Yukawa coupling vanishes.
With the fermionic hopping parameter expansion
it is shown that the fermion mass is proportional
to the Higgs vacuum expectation value
and that the coupling of fermion to the external gauge field
can be chiral if the hopping parameter of the scalar regulator field is taken
to the critical value.
\par}
\vfill
\eject
\chapter

\noindent
{\bf 1. Introduction}

  The non-perturbative formulation of chiral gauge theories is one of the
important problems in quantum field theory, since such a theory describes the
electroweak interactions. The lattice construction of chiral gauge theories
may be one of possible ways toward this goal.
There is, however, an obstacle known as the species doubling
problem\r{\Ninomiya}.

In the approach to chiral gauge model following the idea of
the Wilson-Yukawa
coupling \r{\Swift - \Kashiwa},  
it was found through several numerical \r{\Bock - \BDS}
and analytical \r{\Mean - \KEhop} 
investigations that
the doublers can actually be removed if the Wilson-Yukawa coupling
is strong enough.
In the strong Wilson-Yukawa coupling region, however,
the fermion mass $m_f$ in units of lattice spacing $a$ is not
proportional to the vacuum expectation value $v$ of the Higgs
field. Hence it does not go to zero at the critical point
of the scalar theory where the vacuum expectation value vanishes.
We need to tune the Yukawa coupling $Y$ to obtain a finite fermion mass
in the continuum limit.
The author of ref.\Anna \ examined a modification of the model in which
the scalar field appearing in the Wilson-Yukawa coupling is taken to be
different
from the Higgs field.
The result of numerical simulation for the fermion mass, however,
showed a behavior similar to that of the original model.

In this paper we consider a further modification of the model which can lead to
the relation $m_f=Y v$ in the global limit that the gauge couplings are turned
off.
We impose shift symmetry\r{\Shift} on all the physical fermion fields in the
action
when the Yukawa coupling vanishes.
For this purpose we introduce a gauge {\it singlet} dummy fermion field for
each physical one with chirality opposite to its counterpart.
Using the fermionic hopping parameter expansion,
we evaluate the fermion mass and the coupling to the external gauge field
in this model.

This paper is organized as follows.
In Sec.2, we review the outcome of the lattice model with
the Wilson-Yukawa coupling.
In Sec.3, we examine the original two-scalar
model of ref.\Anna \ 
and clarify the reason why the fermion mass
does not vanish at the critical point.
In Sec.4, we introduce our model and examine the behavior of fermion mass.
In Sec.5, we calculate the coupling of fermion to the external gauge field and
discuss its behavior near the scaling region.
Our conclusion is given in sec. 6.
In the appendix we give the detail of the hopping parameter expansion.

\chapter

\vskip 30pt
\noindent
{\bf 2. Model with Wilson-Yukawa coupling}

We consider the Higgs-fermion part of the standard model as the ``target
theory''.
For simplicity we replace
the $SU(2)_L$ gauge group with an abelian group $U(1)_W$ and neglect
hypercharge. The relevant part of the lagrangian is written as
$$\Lag =-\bar\psi_L \gamma_\mu D_\mu \psi_L
               -\bar{t}_R \gamma_\mu \partial_\mu t_R
               -(D_\mu \phi )^\dagger D_\mu \phi
               -y( \bar\psi_L \phi t_R + \bar{t}_R \phi^\dagger \psi_L)
               -\lambda (\phi^\dagger\phi - v_0^2)^2,
\eqno\eq$$
where $D_\mu \equiv (\partial_\mu+ie W_\mu)$ and
$\phi$ is the $U(1)_W$ Higgs field.
The $U(1)_W$ gauge transformation is defined by
$$\psi_L\arg{x} \longrightarrow e^{i\alpha (x)} \psi_L\arg{x}, \quad
 \phi\arg{x} \longrightarrow e^{i\alpha (x)} \phi\arg{x}, \quad
 W_\mu\arg{x} \longrightarrow W_\mu\arg{x} - {1\over e}\partial_\mu \alpha
(x).
\eqno\eq$$

It is straightforward to discretize the action on the four dimensional
Euclidean lattice. Usually it is defined in the $\lambda\rightarrow\infty$
limit
following the universality hypothesis between the linear and nonlinear $\sigma$
model,
and thereby $\phi\arg{n}$ is replaced by $g\arg{n} \in U(1)$. Such action,
however,
describes the sixteen-fold multiple species of fermions
which appears in the target theory.
In the approach with use of the Wilson-Yukawa coupling
the following lattice action is taken instead of the naively discretized one.
Here we set the lattice constant $a$ equal to one.
$$\eqalign{
S=& \sum_n
\Bigl\{
-{1\over2} \Bigl(
\bar\psi_L\arg{n}\gamma_\mu D_\mu \psi_L\arg{n}
+\bar{t}_R\arg{n}\gamma_\mu \nabla_\mu t_R\arg{n}
\Bigr)
\cr
&\hskip 1cm +{\displaystyle w\over2}
\Bigl(
\bar\psi_L\arg{n} g\arg{n}
\nabla^2 t_R\arg{n}
+ \bar{t}_R\arg{n} \nabla^2 (g^\dagger\psi_L)\arg{n}
 \Bigr)
\Bigr\} \cr
&-\sum_n y \Bigl( \bar\psi_L\arg{n} g\arg{n} t_R\arg{n} +
                  \bar{t}_R\arg{n} g^\dagger\arg{n} \psi_L\arg{n}
           \Bigr)
\cr
&+\sum_{n,\mu} K_H \Bigl( g^\dagger\arg{n} U_\mu\arg{n} g\arg{n+\hat\mu}
           +g^\dagger\arg{n+\hat\mu}U_\mu^\dagger\arg{n} g\arg{n} \Bigr).
\cr
}
\eqno\eq$$
where
$$\nabla_\mu \psi\arg{n}\equiv \psi\arg{n+\hat\mu}-\psi\arg{n-\hat\mu},
\qquad
\nabla^2 \psi\arg{n} \equiv \sum_\mu
 \Bigl( \psi\arg{n+\hat\mu}+\psi\arg{n-\hat\mu}-2\psi\arg{n} \Bigr) .
\$$$
$$
U_\mu\arg{n}\equiv \exp( ieW_\mu\arg{n} ), \qquad
D_\mu \psi\arg{n}\equiv U_\mu\arg{n} \psi\arg{n+\hat\mu}
                       -U_\mu^\dagger\arg{n-\hat\mu}\psi\arg{n-\hat\mu},
\$$$
The second line in the above action is the Wilson-Yukawa term with the
Wilson-Yukawa coupling $w$.
Note that the action is invariant under the global $U(1)_B$
transformation:
$$\psi_L\arg{n} \longrightarrow e^{i\beta} \psi_L\arg{n}, \quad\quad
t_R\arg{n} \longrightarrow e^{i\beta} t_R\arg{n}. \eqno\eq$$
as well as the chiral $U(1)_W$ gauge transformation:
$$
U_\mu\arg{n} \longrightarrow e^{i\alpha\arg{n}} U_\mu\arg{n}
e^{-i\alpha\arg{n+\hat\mu}}, \quad
\psi_L\arg{n} \longrightarrow e^{i\alpha\arg{n}} \psi_L\arg{n}, \quad
g\arg{n} \longrightarrow e^{i\alpha\arg{n}} g \arg{n}. \eqno\eq$$

This model in the global limit
has been examined by numerical
simulations\r{\Bock - \BDS} and analytical methods
\r{\Mean - \KEhop}.
In the region of strong Wilson-Yukawa coupling $r$ and positive hopping
parameter $K_H$ for the Higgs scalar, there are two phases;
the chirally symmetric phase and the chirally broken phase with nonvanishing
vacuum
expectation value of the Higgs scalar.
The species doublers acquire masses of the order of cutoff in
this region.
However the mass of the physical mode in unit of lattice spacing $a$
is not proportional to the vacuum expectation value of the Higgs scalar.
Therefor we need the fine-tuning of the Yukawa coupling to obtain the
finite fermion mass in the continuum limit.

Here we review how
these results for fermion mass can be derived by hopping parameter
expansion\r\Hop.
In terms of the ``neutral'' fermion defined by
$\chi\arg{n} \equiv g^\dagger \arg{n} \psi_L\arg{n}+ t_R\arg{n}$,
the Wilson-Yukawa term in the above action becomes a free Wilson term
$
{\displaystyle r\over2}
\, \bar\chi\arg{n} \nabla^2 \chi\arg{n} ,
$
and the results of the hopping parameter expansion$^{\rm f1}$ with respect to
$K \equiv 1/\arg{8w+2y}$)
agree well
with the results of the numerical
simulations\r{\BDS,\Hop}.
At the leading order the inverse propagator of $\chi$ in the momentum space
is evaluated as
$$
\langle \chi \, \bar\chi \rangle^{-1} \arg{k}
= (4r+Y)\Big( 1+2K
[i(P_R z^2+P_L)\gamma_\mu \sin k_\mu -r\sum_\mu \cos k_\mu ]
 \Big),
\$$$
where $z^2 = \langle g^\dagger\arg{n}g\arg{n+\hat\mu} \rangle$.
The masses of physical mode and doubling modes of $\chi$ are obtained as
$$
m_f a = y / z,
\$$$
$$
m_D a = (y + 2 \, r \, n)/z,
\$$$
where $n$ means the number of momentum components which have the value $\pi$.
Note that the fermion mass $m_f$ is not proportional to the vacuum expectation
value of the Higgs scalar.
The continuum limit is defined at the second order phase transition point
between the symmetric and broken phases,
$$
K_H \longrightarrow K_{H}^c \, ; \quad \langle g \rangle \equiv v a
\longrightarrow 0.
\$$$
In this limit $z$ is known to remain non-vanishing.
Therefore by tuning $y$ as
$
y \longrightarrow 0
$ and only by this tuning
we can make the physical fermion mass finite
in the continuum limit.
The doublers have masses of cutoff order and are decoupled at the
same time.

This behavior of fermion mass
is different from what is expected in the standard model.
If the lattice model would have the relation $m_f = y v$ implied by
the spontaneous breaking of chiral symmetry in the standard model,
the fermion mass in unit of $a$ should vanish at the critical point of the
Higgs scalar without any tuning of the Yukawa coupling.

\chapter

\vskip 30pt
\noindent
{\bf 3. Two-scalar model}

The model considered in ref.\Anna \ contains two scalar fields,
the Higgs field $g\arg{n}$ and another field $w\arg{n}$. The latter
is introduced solely for the purpose of removing fermion doublers and
we refer to it as the Wess-Zumino scalar.
The Wilson-Yukawa coupling contains only $w\arg{n}$,
while the physical Yukawa coupling is introduced among $\psi_L\arg{n}$,
$t_R\arg{n}$ and $g\arg{n}$.
The action in the global limit is given by
$$
S=S_B+S_F,
\$$$
where
$$
S_B= \sum_{n,\mu} K_H \Bigl( g^\dagger\arg{n}g\arg{n+\hat\mu}
+ {\rm h.c.} \Bigr)
+\sum_{n,\mu} K_{wz} \Bigl( w^\dagger\arg{n}w\arg{n+\hat\mu}
+ {\rm h.c.} \Bigr),
\$$$
$$\eqalign{
S_F=&\sum_n \Bigl\{
-{1\over2} \bar{t}_R\arg{n}\gamma_\mu \nabla_\mu t_R\arg{n}
-{1\over2} \bar\psi_L\arg{n}\gamma_\mu \nabla_\mu \psi_L\arg{n}
\cr
&\hskip 3cm +{\displaystyle r\over2}
 \Bigl(
\bar\psi_L\arg{n} w\arg{n}
\nabla^2 t_R\arg{n}
+ {\rm h.c.}
 \Bigr)
\Bigr\}
\cr
&\hskip .5cm -\sum_n Y
\Bigl( \bar\psi_L\arg{n} g\arg{n} t_R\arg{n}
+ {\rm h.c.}
 \Bigr).
\cr
}
\$$$
Since the Wess-Zumino scalar is unphysical it is assumed
to be very heavy in the symmetric phase.
This means that the hopping parameter $K_{wz}$ for $w\arg{n}$
is kept very small.

At the leading order of the hopping parameter expansion in $K\equiv 1/8r$
(see Appendix),
the fermion mass of the $\chi$ field defined by
$\chi\arg{n} \equiv w^\dagger \arg{n} \psi_L\arg{n}+ t_R\arg{n}$
is calculated as
$$
m_f a= Y \Big( \langle w^\dagger\arg{n}g\arg{n} \rangle \Big)/z,
\$$$
where
$z^2 = \langle w^\dagger\arg{n} w\arg{n+\hat\mu} \rangle $ .
To evaluate the expectation value in (3.4) we expand
the fermion determinant up to order $K^2$ and find
$$\eqalign{
\Delta S=& 4KY \sum_n (w^\dagger\arg{n}g\arg{n}
+{\rm h.c.})
+2K^2 \sum_{n,\mu} \Bigl( w^\dagger\arg{n} w\arg{n+\hat\mu}
+ {\rm h.c.} \Bigr)
\cr
&-4K^2 Y^2 \sum_n \Bigl(
 (w^\dagger\arg{n}g\arg{n})^2
+ {\rm h.c.} \Bigr)
 + {\cal O}\arg{K^3}.
\cr
}
\$$$
Using $S_B + \Delta S$ as a total action we then obtain
$$\eqalign{
\langle w^\dagger\arg{n}g\arg{n} \rangle
&=\langle w^\dagger\arg{n}\rangle_0  \langle g\arg{n} \rangle_0
+4KY+{\cal O}(K^2)
\cr
&=4KY+{\cal O}(K^2).
\cr
}
\$$$
where $\langle \quad \rangle_0$ means the expectation value with respect to
$S_B$.
The second equality holds since $K_{wz}$ is smaller than its critical value and
hence $\langle w^\dagger\arg{n}\rangle_0 = 0$.
Note that the first and the third term on the right hand side of (3.5)
give rise to a mixing between $g$ and $w$ and this is the origin of the
contribution $4KY$ in (3.6).
{}From (3.4) and (3.6) we obtain
$$
m_f a = 4KY^2 /z_0 + {\cal O}(K^2).
\$$$
where
$ z_0^2 = \langle w^\dagger\arg{n} w\arg{n+\hat\mu} \rangle_0 $.
This result explains the numerical data of ref.\Anna \ that
the fermion mass does not go to zero as $K_H$ tends to the critical value.
This is because
due to the strong Wilson-Yukawa coupling
the field $\chi$ with a free Wilson term becomes physical,
its mass is proportional to
$\langle w^\dagger g \rangle$ instead of
$\langle  g \rangle$ and the Yukawa coupling
induces the mixing between $g$ and $w$ leading to
nonvanishing $\langle w^\dagger g \rangle$.
Therefore $Y$ has to be tuned in order to obtain finite $m_f$
in the continuum limit.
Our calculation also explains the observed fact
that $\langle w \rangle$ is proportional to $\langle g \rangle$\r{\Anna}.
Namely
$$
\langle w\arg{n} \rangle = 4KY \langle g\arg{n} \rangle + {\cal O}(K^2).
\$$$

One may add local counterterms, which are similar to those discussed in the
Rome
approach\r{\Rome}, in order to remove the mixing
between $\omega$ and $g$ in $\Delta S$.
This implies, however, that the mass of $\chi$ is identically zero
irrespective of the value of the Yukawa coupling $Y$.
Therefore the relation $m_f=Yv$ can not be obtained with such a subtraction.

\chapter

\vskip 30pt
\noindent
{\bf 4. Physical fields with shift symmetry and $m_f=Y v$ relation}

Now we consider
a further modification of the model which can lead to
the relation $m_f=Y v$ in the global limit.
The model contains gauge-singlet dummy fields $\psi_R$ and $t_L$
of opposite chirality to the physical fermion fields $\psi_L$ and
$t_R$\r{\Rome,\Ki}.
The Yukawa interaction is introduced only among the physical fields
$\psi_L$, $t_R$ and $g$.
The Wilson-Yukawa term is constructed between the corresponding physical
and dummy fields so that
the ``charged'' fermion field $\psi^c\arg{n} \equiv
\psi_L\arg{n}+w\arg{n}\psi_R\arg{n}
$ and the field $t\arg{n}$ have free Wilson-terms.
The action is given by
$$\eqalign{S_F =& \sum_n \Bigl\{
-{1\over2} \bar{t}\arg{n} \gamma_\mu \nabla_\mu t\arg{n}
+{r\over 2} \, \bar{t}\arg{n} \nabla^2 t\arg{n}
\cr
&
-{1\over2}
\bar\psi\arg{n} \gamma_\mu \nabla_\mu \psi\arg{n}
+{r\over 2} \Bigl(
\bar\psi_L\arg{n}\nabla^2
 w\arg{n} \psi_R\arg{n}
+{\rm h.c.} \Bigr)\, \Bigr\}
\cr
&\hskip .5cm -\sum_n Y \Bigl( \bar\psi_L\arg{n} g\arg{n} t_R\arg{n}
+{\rm h.c.}        \Bigr)
\cr
}
\$$$
The chiral symmetry $U(1)_W$ which will be gauged is defined as
$$
\psi_L \rightarrow e^{i\alpha} \psi_L, \quad
g \rightarrow e^{i\alpha} g , \quad
w \rightarrow e^{i\alpha} w .
\$$$
The model has a vector-like global symmetry $U(1)_B^\prime$
$$
\psi \rightarrow e^{i\beta} \psi, \quad
t \rightarrow e^{i\beta} t,
\$$$
and an additional global symmetry $U(1)_R$:
$$
\psi_R \rightarrow e^{i\phi} \psi_R, \quad
 w \rightarrow e^{-i\phi} w.
\$$$

The action above has a shift symmetry\r{\Shift} under
$t_L \arg{n} \rightarrow t_L\arg{n} +\epsilon^\prime_L
$ for non-zero $Y$.
In our choice of the Wilson-Yukawa coupling, further
shift symmetry for physical fields emerges at $Y=0$:
$\psi_L\arg{n} \rightarrow \psi_L\arg{n} + \epsilon_L$, and
$t_R \arg{n} \rightarrow t_R\arg{n} +\epsilon_R$.
This symmetry ensures that
the charged fermion field and the field $t\arg{n}$
appear as massless particles at $Y=0$\r{\Shift}.
With the hopping parameter expansion in terms of these fields
the fermion mass connecting $\psi_L$ and $t_R$ is obtained as
$$
m_f a = Y \langle g\arg{n} \rangle + {\cal O}(K^2).
\$$$
The desired relation $m_f a = Y \langle g \rangle$ is thus derived in this
model.
In the limit that $\langle g \rangle \rightarrow 0$, where
the physical fermion mass $m_f$ remains finite,
the masses of doublers behave as
$$
M_D(\psi^\prime)a \longrightarrow 2rn/z,
\$$$
$$
M_D(t^\prime)a \longrightarrow 2rn .
\$$$
The doublers are all heavy and decouples in this continuum limit.
Note also that the fermion determinant up to order ${\cal O}(K^2)$ takes the
form
$$
\Delta S =
2K^2 \sum_{n,\mu} \Bigl( w^\dagger\arg{n}w\arg{n+\hat\mu}
+{\rm h.c.} \Bigr),
\$$$
and $w\arg{n}$ decouples from $g\arg{n}$ without any mixing. Its effect
is only to shift the hopping parameter $K_{wz}$,
$$
K_{wz} \rightarrow K^\prime_{wz} \equiv K_{wz} + 2K^2.
\$$$

\chapter

\vskip 30pt
\noindent
{\bf 5. Coupling to external gauge field}

If one set $Y$ equal to zero, the part of the $\psi\arg{n}$ field
in  eq. (4.1) reduces to the model discussed in ref.\GPS.
It has been argued that the charged fermion has the vector-like coupling
with the external gauge field in the symmetric phase.
On the other hand, the calculation by the hopping parameter expansion
in ref.\Aokif \ shows
that the fermion can have chirally asymmetric interactions.
In fact the three point vertex among the charged fermion $\psi^c\arg{n}$
and an external gauge field $A_{\mu}\arg{l}$ is given in the leading order
of hopping parameter expansion by
$$\eqalign{
\Gamma_\mu\arg{n,m,l} =&{1\over 2}
 \Big\{ [\gamma_\mu P_L-r]\delta_{n+\hat\mu,m}\delta_{n,l}
+( n \leftrightarrow m)
\Big\}
\cr
&-K_{wz} \gamma_\nu P_R
 \Big\{
  \langle Z_\mu\arg{l} Z_\nu\arg{n} \rangle_c \delta_{n+\hat\nu,m}
+(n \leftrightarrow m) \Bigr\}
\cr
&+{\cal O}\arg{K},
\cr
}
\$$$
where $Z_\mu\arg{n}=Im ( w\arg{n}w^\dagger\arg{n+\hat\mu})$
through which
the coupling of $\psi^c_R$ to $A_{\mu}$ takes place.
Thus the coupling of fermion with the external field is neither completely
chiral nor vector-like
before the continuum limit is taken.
There are two possible continuum limits for this model.
First we consider the case where
the Wess-Zumino scalar $w$ becomes infinitely heavy.
For the case, we only have to keep the
inequality $K_{wz} < K_{wz}^{\rm critical}$.
No fine-tuning of $K_{wz}$ is needed.
We then have
$$
 \langle Z_\mu\arg{l} Z_\nu\arg{n} \rangle_c
\rightarrow   \delta_{l,n}\delta_{\mu,\nu}
\simeq a^4\delta (la-na)\delta_{\mu,\nu} ,
\$$$
and therefore
$$\eqalign{
\Gamma_\mu\arg{n,m,l} \rightarrow &{1\over 2}
 \Big\{ [\gamma_\mu (P_L+P_R K_{wz})-r]\delta_{n+\hat\mu,m}\delta_{n,l}
\cr
&\hskip 1cm +[\gamma_\mu (P_L+P_R K_{wz})+r]\delta_{n,m+\hat\mu}\delta_{m,l}
\Big\}.
\cr
}
\$$$
In the same limit the inverse propagator of $\psi^c$ is given by
$$\eqalign{
S^{-1}\arg{n,m} \rightarrow&
{1\over 2}\sum_\mu
 \Big( [\gamma_\mu (P_L+P_R K_{wz})-r]\delta_{n,m+\hat\mu}
\cr
&\hskip 1cm -[\gamma_\mu (P_L+P_R K_{wz})+r]\delta_{n+\hat\mu,m} +2r
\delta_{n,m}\Big),
\cr
}\$$$
and it determines the wave function renormalization factor of the fermion:
$$
Z_2=(P_L+P_R K_{wz}^{-1}).
\$$$
The renormalization factor cancels $K_{wz}$ of eq.(5.3) in the renormalized
vertex
$\Gamma_\mu^{\rm REN}$
and we find a vector-like coupling:
$$
\Gamma_\mu^{\rm REN}(x,y,z) \rightarrow
\delta (x-y)\delta (x-z) \, \gamma_\mu .
\$$$
Keeping $w$ infinitely heavy in the continuum limit
leads to a theory with vector-like coupling\r{\GPS}.

The second possible continuum limit is given
by tuning$^{\rm f2}$ $K_{wz}$ to its criticality
$K_{wz} \rightarrow K_{wz}^{\rm critical}$.
In this case we expect the mass of the Wess-Zumino scalar to be finite.
Denoting by $\phi$ the renormalized field operator for the Wess-Zumino scalar
in the continuum limit, $Z_\mu$ can be expressed as
$$\eqalign{
Z_\mu =& a^3 [z_a i (\phi^\dagger \partial_\mu \phi
                      -\partial_\mu(\phi^\dagger) \phi )
             +z_b \partial_\mu (\phi^\dagger \phi)] \dots
\cr
\equiv& a^3 Z_\mu^{\rm REN} +\dots,
\cr
}
\$$$
on dimensional grounds,
where $z_a$ and $z_b$ give the overlapping between $Z_\mu$ and the
corresponding operators, respectively.
We then obtain
$$
 \langle Z_\mu\arg{l} Z_\nu\arg{n} \rangle_c
\rightarrow
a^6 \,
\langle Z_\mu^{\rm REN}\arg{l} Z_\nu^{\rm REN}\arg{n} \rangle_c.
\$$$
Even when the renormalized two point function above contains the
local contact term proportional to the delta function, it
leaves the factor of $a^2$. Therefore we have
$$
\Gamma_\mu^{\rm REN} \rightarrow
\delta (x-y)\delta (x-z)\gamma_\mu (P_L + O(a^2)P_R) ,
\$$$
and chiral coupling can be obtained in this limit.
Even though the unphysical Wess-Zumino scalar appears at the physical scale,
there is a possibility to keep it from interacting with other particles
such as the gauge and Higgs particles.
This condition corresponds to the recovery of the gauge(BRST) invariance
of the target theory as discussed in ref.\Rome.
It requires further fine-tuning of parameters besides the tuning of $K_{wz}$.

\chapter

\vskip 30pt
\noindent
{\bf 6. Discussion and Conclusion }

We have to mention about the anomaly of the baryon number current.
In the model discussed here the baryon number is broken explicitly by
the Wilson-Yukawa term. If we could add successfully the
counterterms to decouple the Wess-Zumino scalar, that is, we would have the
BRST invariance of the target theory, the current should have a correct
anomaly.
As far as we restrict ourselves in the perturbative treatment of the
gauge field, this program works just like the continuum theory.
However we do not yet understand whether the nonperturbative program would
really work.

As a conclusion, we have shown that it is possible to formulate
a chiral Yukawa model
which satisfies the relation $m_f=Yv$ while doublers are removed by
the Wilson-Yukawa coupling. This is achieved through the introduction
of an additional scalar field and gauge-singlet dummy fermion fields having
opposite chirality to the physical fermions. By virtue of these fields
all the physical fermion fields can be made to possess shift symmetry when
the Yukawa coupling vanishes.
It is also argued that the coupling of fermion to the external gauge
field can be chiral at the critical point of the additional scalar.
Further fine-tuning of parameters are, however, needed for the decoupling
of the additional scalar from the gauge and Higgs particles and for the
gauge(BRST) invariance of the target theory.
Nonperturbative prescription for the gauge invariance in this formulation
is not yet understood.

\vskip 30pt
\noindent
{\bf Acknowledgement}

We would like to thank Y. Iwasaki, T. Hattori, K. Kanaya, T. Kashiwa,
K-I. Kondo, M. Okawa, R.E. Shrock and A. Ukawa for valuable
discussions. Y.K. is supported by a Grant-in-Aid for
JSPS fellow.

\vskip 30pt
\Appendix
\chapter

\noindent
{\bf Appendix}

\vskip 10pt
In this appendix, we give the detail of the hopping
parameter expansion omitted in the main text.

If the fermionic action is written as
$$
S_F =- \sum_{n,m}
\bar\Psi \arg{n} C\Big( 1_{n,m}+K D_{n,m} \Big) \Psi\arg{m},
\eqno\eq$$
with some small parameter $K << 1$ ($C$ is a normalization
constant),
the fermion determinant and the propagator can be expanded in
terms of K.
$$\eqalign{
\Delta S&= \Tr \Ln  \Big( 1_{n,m}+K D_{n,m} \Big)  \cr
        &=          \Tr \Big( K D_{n,m} \Big)
          -{1\over2}\Tr \Big( K D_{n,m} \Big)^2
          +{\cal O}(K^3) + \cdots, \cr
}
\eqno\eq$$
$$\eqalign{
\Big\langle \Psi\arg{n} \bar\Psi\arg{m} \Big\rangle
&=\Big\langle \Bigl\{ C \Big( 1_{n,m}+K D_{n,m} \Big) \Bigr\}^{-1} \Big\rangle
\cr
&= {\displaystyle 1 \over C}
\Big( 1_{n,m}- K \Big\langle D_{n,m} \Big\rangle
+ K^2 \Big\langle D^2_{n,m} \Big\rangle  + {\cal O}(K^3) + \cdots \Big),
\cr
}
\eqno\eq$$
where $\langle \ \rangle$ means the expectation value evaluated with
the original bosonic action and the fermion determinant;
$$
\langle {\cal O} \rangle \equiv
\int [d\phi d\phi^\dagger]\; {\cal O} \;
e^{S_B[\phi]+\Delta S[\phi]} \, / Z.
\eqno\eq$$
This expectation value can be evaluated in the series expansion of $K$
since the fermion determinant is also evaluated in the expansion.

In the following, $d$ means the number of space-time dimension.

\vskip 10pt
\noindent
{\bf The two scalar model in Sec. 3}

In the two scalar model in Sec. 3, we define ``neutral'' field
as follows.
$$\chi\arg{n} \equiv w^\dagger \arg{n} \psi_L\arg{n}+ t_R\arg{n}
.\eqno\eq$$
Then the fermionic action is written as
$$
S_F=\sum_{n,m}
\bar\chi\arg{n} (-rd)\Big( 1_{n,m}+K D_{n,m} \Big) \chi\arg{m},
\eqno\eq$$
where
$$\eqalign{
K&={1\over 2rd} \; ,
\cr
D_{n,m}
&= 2Y(P_R w^\dagger\arg{n}g\arg{n}+P_L g^\dagger\arg{n}w\arg{n}) \delta_{n,m}
\cr
&-\sum_\mu \Bigl\{
\bigl[ r-\gamma_\mu(P_Lw^\dagger\arg{n}w\arg{m}+P_R )\bigr]
\delta_{n+\hat\mu,m}
+\bigl[ r+\gamma_\mu(P_Lw^\dagger\arg{n}w\arg{m}
+P_R )\bigr]
\delta_{n,m+\hat\mu}
\Bigr\}.
\cr
}
\eqno\eq$$

The fermion determinant is evaluated in the expansion with respect to $K
(<<1)$.
$$
\Delta S = \Tr \Big( K D_{n,m} \Big)
          -{1\over2}\Tr \Big( K D_{n,m} \Big)^2
          +{\cal O}(K^3),
\eqno\eq$$
where
$$\eqalign{
\Tr \Big( K D_{n,m} \Big)
&=4KY \sum_n (w^\dagger\arg{n}g\arg{n}+g^\dagger\arg{n}w\arg{n}),
\cr
-{1\over2}\Tr \Big( K D_{n,m} \Big)^2
&=2 K^2 \;
\sum_{n,\mu}( w^\dagger\arg{n}w\arg{n+\hat\mu}+w^\dagger\arg{n+\hat\mu}w\arg{n}
)
\cr
&\hskip 1cm - 4 K^2Y^2
\sum_n \Big(
(w^\dagger\arg{n}g\arg{n})^2+(g^\dagger\arg{n}w\arg{n})^2
       \Big) \quad + {\rm const.}\; .
\cr
}
\eqno\eq$$

The fermion propagator is also evaluated as
$$\eqalign{
\Big\langle \chi\arg{n} \bar\chi\arg{m} \Big\rangle
&=\Big\langle \Bigl\{ (rd)\Big( 1_{n,m}+K D_{n,m} \Big) \Bigr\}^{-1}
\Big\rangle
\cr
&= {\displaystyle 1 \over rd}
\Big( 1_{n,m}- K \Big\langle D_{n,m} \Big\rangle  +{\cal O}(K^2) \Big),
\cr
}
\eqno\eq$$
and its inverse is given by
$$\eqalign{
& \Big\langle \chi\arg{n} \bar\chi\arg{m} \Big\rangle ^{-1}
= rd \, \delta_{n,m}
+Y \Big( P_R \langle w^\dagger\arg{n}g\arg{n} \rangle
  +P_L \langle g^\dagger\arg{n} w\arg{n} \rangle\Big) \delta_{n,m}
\cr
&\hskip 3.0cm -{1\over2} \sum_\mu \Big\{
\Big[ r
 -\gamma_\mu \Big( P_L\langle w^\dagger\arg{n}w\arg{m}\rangle+P_R \Big)
\Big]
\delta_{n+\hat\mu,m}
\cr
&\hskip 5.5cm
+\Big[ r
+\gamma_\mu \Big( P_L\langle w^\dagger\arg{n}w\arg{m}\rangle+P_R \Big)
\Big]
\delta_{n,m+\hat\mu}
\Big\}
\cr
& \hskip 3.cm
-2KY^2 \Big( P_R \langle (w^\dagger\arg{n}g\arg{n})^2
                                               \rangle_{c^\prime}
  +P_L \langle (g^\dagger\arg{n} w\arg{n})^2 \rangle_{c^\prime}
\Big) \delta_{n,m}
\cr
&-KY \sum_\mu \Big\{
\gamma_\mu P_L
\langle
[ w^\dagger\arg{n}g\arg{n}+g^\dagger\arg{n+\hat\mu}w\arg{n+\hat\mu}]
              w^\dagger\arg{n}w\arg{n+\hat\mu} \rangle_{c^\prime}
\delta_{n+\hat\mu,m}
\cr
&\hskip 2.cm
-\gamma_\mu P_L
\langle
[ w^\dagger\arg{m+\hat\mu}g\arg{m+\hat\mu}+g^\dagger\arg{m}w\arg{m}]
              w^\dagger\arg{m+\hat\mu}w\arg{m} \rangle_{c^\prime}
\delta_{n,m+\hat\mu}
\Big\}
\cr
&+{\cal O}(K^2) ,
\cr
 }
\eqno\eq$$
where
$$
\langle w^\dagger\arg{n} g\arg{n} \rangle
=\langle w^\dagger\arg{n} \rangle_0 \, \langle g\arg{n} \rangle_0 + 4KY
+{\cal O}(K^2),
\eqno\eq$$
and $\langle \quad \rangle_{c^\prime}$ means the connected parts defined as
follows:
$$\eqalign{
\langle (w^\dagger\arg{n}g\arg{n})^2 \rangle_{c^\prime}
&\equiv
\langle (w^\dagger\arg{n}g\arg{n})^2 \rangle
-
\langle w^\dagger\arg{n}g\arg{n} \rangle^2,
\cr
\langle
( w^\dagger\arg{n}g\arg{n}+
g^\dagger
& \hskip .05cm \arg{n+\hat\mu}w\arg{n+\hat\mu})
 w^\dagger\arg{n}w\arg{n+\hat\mu} \rangle_{c^\prime} \cr
&\equiv
\langle
( w^\dagger\arg{n}g\arg{n}+g^\dagger\arg{n+\hat\mu}w\arg{n+\hat\mu})
              w^\dagger\arg{n}w\arg{n+\hat\mu} \rangle            \cr
&\hskip 16pt
-\langle
( w^\dagger\arg{n}g\arg{n}+g^\dagger\arg{n+\hat\mu}w\arg{n+\hat\mu})
  \rangle
  \langle w^\dagger\arg{n}w\arg{n+\hat\mu} \rangle.  \cr
}
$$
We also obtain the following relation.
$$
\langle w\arg{n} \rangle = \langle w\arg{n} \rangle_0
+  4KY \, \langle g\arg{n} \rangle +{\cal O}(K^2).
\eqno\eq$$

\vskip 10pt
\noindent
{\bf The model in Sec. 4}

For the model in Sec. 4, we define a two-component field as
$$\Psi\!^c \arg{n} \equiv  \left(
 \matrix{ t\arg{n}                                       \cr
          \psi_L\arg{n}+ w\arg{n} \psi_R\arg{n}    \cr} \right)
=\left(\matrix{ t\arg{n}                          \cr
                \psi^c\arg{n}  \cr } \right)
.\eqno\eq$$

\noindent
Then the fermionic action can be written as
$$\eqalign{
S_F
&=\sum_{n,m}
\bar\Psi^c \; \arg{n} (-rd)\Big( 1_{n,m}+K {\cal D}^c_{n,m} \Big)
\Psi\!^c\arg{m} ,
\cr
}
\eqno\eq$$
where
$$\eqalign{
K&={1\over 2rd} \; ,
\cr
{\cal D}^c_{n,m}
&= 2Y(\tau^{-} P_R g\arg{n} + \tau^{+} P_L g^\dagger\arg{n}) \delta_{n,m}
\cr
&-\Big({\displaystyle 1+\tau^3 \over 2}\Big) \sum_\mu \Bigl\{
 [ r-\gamma_\mu ]\delta_{n+\hat\mu,m}+[ r+\gamma_\mu ] \delta_{n,m+\hat\mu}
\Bigr\}
\cr
&-\Big({\displaystyle 1-\tau^3 \over 2}\Big)\sum_\mu \Bigl\{
\bigl[ r-\gamma_\mu(P_L+P_R w\arg{n}w^\dagger\arg{m})\bigr]
\delta_{n+\hat\mu,m}
\cr
&\hskip 5cm
+\bigl[ r+\gamma_\mu(P_L+P_R w\arg{n}w^\dagger\arg{m})\bigr]
\delta_{n,m+\hat\mu}
\Bigr\},
\cr
}
\eqno\eq$$
$$
\tau^i \; ; \; {\rm Pauli\ matrices}, \quad \tau^{-}={\displaystyle
\tau^1-i\tau^2 \over2}, \;
\tau^{+}={\displaystyle \tau^1+i\tau^2 \over2}.
\eqno\eq$$

The fermion determinant can be evaluated as follows.
$$
\Delta S
=2 K^2  \;
\sum_{n,\mu}( w^\dagger\arg{n}w\arg{n+\hat\mu}+w^\dagger\arg{n+\hat\mu}w\arg{n}
)
+{\cal O}(K^4) .
\eqno\eq$$
The inverse fermion propagator is evaluated as
$$\eqalign{
\Big\langle \Psi^c\arg{n} \bar\Psi^c\arg{m} \Big\rangle ^{-1}
&= rd \,
\Big( 1_{n,m}+ K \Big\langle {\cal D}^c_{n,m} \Big\rangle  +{\cal O}(K^2) \Big)
\cr
&= Y(\tau^{-} P_R \langle g\arg{n} \rangle
    + \tau^{+} P_L \langle g^\dagger\arg{n} \rangle ) \delta_{n,m}
\cr
&+ rd \, \delta_{n,m}
\cr
&-\Big({\displaystyle 1+\tau^3 \over 2}\Big) {1\over2}\sum_\mu \Bigl\{
 [ r-\gamma_\mu ]\delta_{n+\hat\mu,m}+[ r+\gamma_\mu ] \delta_{n,m+\hat\mu}
\Bigr\}
\cr
&-\Big({\displaystyle 1-\tau^3 \over 2}\Big) {1\over2}\sum_\mu \Bigl\{
\bigl[ r
       -\gamma_\mu(P_L+P_R \langle w\arg{n}w^\dagger\arg{m} \rangle )\bigr]
\delta_{n+\hat\mu,m}
\cr
&\hskip 5cm
+\bigl[ r
+\gamma_\mu(P_L+P_R \langle w\arg{n}w^\dagger\arg{m} \rangle )\bigr]
\delta_{n,m+\hat\mu}
\Bigr\}
\cr
&+{\cal O}(K),
\cr
 }
\eqno\eq$$
where
$$
\langle w\arg{n}w^\dagger\arg{n+\hat\mu}\rangle
=\langle w\arg{n}w^\dagger\arg{n+\hat\mu}\rangle_0 + {\cal O}(K^2),
\eqno\eq$$
$$
\langle g\arg{n} \rangle
=\langle g\arg{n} \rangle_0 .
\eqno\eq$$
{}From this result we can extract the mass terms of physical fermion and
doublers.
For the physical fermion we obtain
$$   - Y \langle g \rangle \bar\psi_L \, t_R \quad + h.c. .
\$$$
For doublers we obtain the mass matrix as
$$
-\left( \matrix{ \bar{t}_L & \bar\psi_L \cr } \right)
\;
\left( \matrix{ 2rn  & Y \langle g \rangle \cr
                0   & 2rn/z                       \cr } \right)
\;\left( \matrix{ t_R \cr
                  \psi_R^{(r)} \cr } \right) ,
\$$$
where $\psi_R^{(r)}$ is already normalized with the wave function
renormalization constant
$Z_2[\psi_R] \equiv
1/z=\langle w\arg{n}w^\dagger\arg{n+\hat\mu}\rangle_0^{-{1\over2}}$.
$n$ means the number of momentum components which have the value $\pi$.
This matrix is diagonalized by changing the basis as
$$
\left( \matrix{ t_R \cr
                  \psi_R^{(r)} \cr } \right) \rightarrow
\left( \matrix{ t_R^\prime \cr
                  \psi_R^{(r)\prime} \cr } \right)
\equiv \left( \matrix{ \cos\theta & \sin\theta \cr
                       -\sin\theta & \cos\theta \cr } \right) \;
       \left( \matrix{ t_R \cr
                  \psi_R^{(r)} \cr } \right),
\$$$
$$
\left( \matrix{ t_L \cr
                  \psi_L \cr } \right) \rightarrow
\left( \matrix{ t_L^\prime \cr
                  \psi_L^\prime \cr } \right)
\equiv \left( \matrix{ \cos\Theta & \sin\Theta \cr
                       -\sin\Theta & \cos\Theta \cr } \right) \;
       \left( \matrix{ t_L \cr
                     \psi_L \cr } \right),
\$$$
where
$$
{1\over2} \tan 2\theta =
{\displaystyle (2rn) (Y \langle g \rangle )
\over
(1/z^2-1)(2rn)^2+(Y \langle g \rangle )^2 },
\$$$
and
$$
\tan\Theta={1\over z} ( \tan\theta + {Y \langle g \rangle a \over 2rn}).
\$$$
The diagonalized masses of doublers are
$$
M_D(t^\prime)a = Cc(2rn)+Ss(2rn/z)-Cs(Y \langle g \rangle )
\buildrel \langle g \rangle  \rightarrow 0 \over \longrightarrow 2rn,
\$$$
$$
M_D(\psi^\prime)a = Cc(2rn/z)+Ss(2rn)+Sc(Y \langle g \rangle )
\buildrel \langle g \rangle  \rightarrow 0 \over \longrightarrow 2rn/z,
\$$$
where we use the abbreviation as
$$
c=\cos\theta, \quad s=\sin\theta,\qquad C=\cos\Theta, \quad S=\sin\Theta.
\$$$

\vfill
\eject
\vskip 40pt
\noindent
{\bf Footnotes}
\chop On the other hand, the hopping parameter expansion
with respect to the ``charged'' fermion defined by
$\psi^c\arg{n}$$\equiv$$\psi_L\arg{n}+g\arg{n}t_R\arg{n}$
$=g \arg{n} \chi\arg{n}$,
with the Wilson-Yukawa term
${r\over2}\bar\psi^c\arg{n} g\arg{n} \nabla^2 (g^\dagger\arg{n}\phi^c\arg{n})$,
is known to perform poorly\r\BDS.
Therefore throughout this paper we use the hopping parameter expansion
with respect to such a field with the free Wilson term.

\qkh $K_{wz}$ is simply replaced by $K^\prime_{wz}$ at the order of $K^2$
since, as shown in sec. 4,
quantum effect does not induce the coupling of $w\arg{n}$ to
$g\arg{n}$ and it only shifts the hopping parameter $K_{wz}$
to $K^\prime_{wz}$.

\vskip 40pt
\noindent
{\bf References}

\ninomiya H.B. Nielsen and M. Ninomiya, \NPB{185}{1981}{20};
\item{ } \NPB{193}{1981}{173} .

\wilson K.G. Wilson, in {\it New phenomena in subnuclear physics},
\item{ } ed. A. Zichichi (Erice 1975) (Plenum, New York, 1977).

\swift P.D.V. Swift, \PLB{145}{1984}{256}.

\smit  J. Smit, Acta Phys. Polon. {\bf B17} (1986) 531.

\aoki   S. Aoki, \PRL{60}{1988}{2109}; $\,$
\PRD{38}{1988}{618}.

\kashiwa  K. Funakubo and T. Kashiwa, \PRL{60}{1988}{2113};
\item{ }   \PRD{38}{1988}{2602}.

\bock W. Bock, A.K. De, K. Jansen, J. Jer\'ak, T. Neuhaus and J. Smit,
\item{ }  \PLB{232}{1989}{486}; \NPB{344}{1990}{207}.

\alss S. Aoki, I-H. Lee, J. Shigemitsu and R. Shrock, \PLB{243}{1990}{403}.

\jansen W. Bock, A.K. De, C. Frick, K. Jansen and T. Trappenberg,
\item{ } \NPB{371}{1992}{683}.

\bds W. Bock, A.K. De and J. Smit, preprint ITFA-91-30,HLRZ-91-81.

\mean J. Smit, \NPBsupl{9}{1989}{579}.

\ivsd M.F.L. Golterman and D.N. Petcher, \PLB{247}{1990}{370}.

\hop S. Aoki, I-H. Lee, and R.E. Shrock, \NPB{355}{1991}{383};
\item{ } \NPBsupl{20}{1991}{589}.

\kehop T. Ebihara and K-I. Kondo, preprint CHIBA-EP-61(1992).

\anna  A. Hasenfratz, \NPBsupl{20}{1991}{555}.

\shift M.F.L. Golterman and D.N. Petcher, \PLB{225}{1990}{159}.

\rome A. Borrelli, L. Maiani, G.C. Rossi, R. Sisto and M. Testa,
\item{ } \NPB{333}{1990}{335}; L. Maiani, G.C. Rossi and M. Testa,
\item{ } \PLB{261}{1991}{479}.

\ki Y. Kikukawa, \MPLA{7}{1992}{871}.

\gps  M. Golterman, D. Petcher and J. Smit, \NPB{370}{1992}{51}.

\aokif S. Aoki, \PROG{88}{1992}{455}.

\vfill
\eject

\bye